\documentclass[preprint,aps,prc,showpacs,nofootinbib]{revtex4}
\usepackage{epsfig}
\usepackage{graphicx,amsmath}

\newcommand\ba{\begin{eqnarray}}
\newcommand\ea{\end{eqnarray}}
\newcommand\be{\begin{equation}}
\newcommand\ee{\end{equation}}
\newcommand\nn{\nonumber}

\begin{document}
\title{Sum rule for the double virtual Compton scattering}
\author{E.~A.~Kuraev}
\affiliation{\it JINR-BLTP, 141980 Dubna, Moscow region, Russian Federation}
\author{S.~Bakmaev}
\affiliation{\it JINR-BLTP, 141980 Dubna, Moscow region, Russian Federation}
\author{V.~V.~Bytev}
\affiliation{\it JINR-BLTP, 141980 Dubna, Moscow region, Russian Federation}

\author{E.~Tomasi-Gustafsson}
\affiliation{\it DAPNIA/SPhN, CEA/Saclay, 91191 Gif-sur-Yvette
Cedex, France }

\date{\today}

\begin{abstract}
The two photon exchange amplitude is investigated in frame of analytic properties of the virtual Compton scattering amplitude as
a function of the invariant mass squared of the intermediate hadronic state. A sum rule is built, based on arguments from analyticity. It relates the differential cross section of elastic
electron-proton scattering including form factors,  and the cross section of inelastic scattering channel, with a contribution of nucleon anti-nucleon pair
production arising from the Fermi statistics. The last term is
calculated in frame of a simple model of nucleon-pion interaction.
\end{abstract}

\maketitle
\section{Introduction}
In a previous paper of one of us \cite{baier81} a sum rule relating the electrons (Dirac and Pauli) form factors
with the inelastic cross sections of processes in electron-positron high energy peripheral (small
angle) scattering was investigated. For $e^+e^-$ peripheral scattering, in the lowest order of perturbation theory (PT), a relation between
the electron Dirac form factor (as a function of transferred momentum square) and the derivative over the transferred momentum square of the cross section for the emission of real photon (soft and hard) was obtained. In the next order of QED PT these sum rules becomes more
complicated. They relate the radiative corrected cross section of a single photon emission, the cross section of two real photon emission, the Dirac and Pauli form factors of electron,
computed in the relevant order of PT, with that contribution to the cross section of $e^+e^-$ pair production which takes into account the identity of two electrons in the final state. The calculation
was given in the logarithmic approximation (Weizs\"acker-Williams approximation), which corresponds to
the limiting case of small values of the transferred momentum. A relation between the total cross section of inelastic processes in peripheral $e^+e^-$ high energy collisions and the slope
of Dirac form factor at zero momentum transfer was obtained.

In Ref. \cite{dubn}, analytical properties of the Compton scattering amplitude which are the basis of these relations, were used to describe the proton block of
the electron-proton scattering amplitude. In this way a relation between the radius and the anomalous
magnetic momentum of proton and neutron, on one side, and the total cross sections of photoproduction on the other side, was derived. The difficulties related with the possible Pomeron contribution to
the description of the photoproduction cross section was overcome by building linear combinations
of cross sections, which are free from Pomeron pole contribution.

Recently,  the relevance of the contribution of the two photon exchange
amplitude (TPE) to the differential cross section was discussed in the literature \cite{twogamma}, in order to solve the discrepancy between new experimental data on elastic electron proton scattering \cite{Jo00}, based on the polarization method \cite{Re68}, and all unpolarized data based on the Rosenbluth fit \cite{Ro50}. Although in a previous work devoted to
this problem \cite{By07}, it was shown that TPE contribution is too small (it does  not exceed $2\%$ and it is not the solution to this problem), the investigation of TPE amplitude is interesting by itself, as it can help
to understand the properties of the Compton scattering amplitude in the low energy region. The motivation of this paper is to study analytical properties of TPE amplitude.
The paper is organized as follows. In Section II, a sum rule is
derived in terms of the discontinuity of scattering amplitudes and of the corresponding contribution to the differential cross section as a function of  the transferred momentum. Section III is devoted to the explicit
calculation of the left cut contribution. The results are discussed in Conclusion. The kinematics of the two-loop Feynman integral, which enters in the calculation of the left cut, is given in the Appendix.

\section{Peripheral kinematics. Derivation of the Sum rule.}
Let us consider elastic electron proton scattering
\be
e(p_1)+p(p)\to e(p_1')+p(p'),
\label{eq:eq1}
\ee
in peripheral kinematics:
\be
q=p_1-p_1',~s=2p_1p>>Q^2=-q^2, p^2=p^{'2}=M^2.
\label{eq:eq2}
\ee
The Born amplitude has the form:
\be
M_B=\frac{4\pi\alpha}{q^2}\bar{u}(p_1')\gamma_\mu u(p_1)\bar{u}(p')\Gamma_\nu u(p) g^{\mu\nu}, \nn
\ee
where the electromagnetic vertex of proton is
\be
\Gamma_\mu(q)=\left [F_1(q^2)+\frac{\hat{q}}{2M}F_2(q^2)\right ]\gamma_\mu,
\ee
and $F_{1,2}$ are the Dirac and Pauli form factors of proton.

The main property of the peripheral amplitude of charged particles scattering is related to non-vanishing differential cross section in the high energy limit. This can be more explicitly  seen
using the Gribov's representation of the photon Green function:
\be
g^{\mu\nu}\approx \frac{2}{s}p^\mu p_1^\nu.
\label{eq:eqa}
\ee
With such substitution we obtain
\be
M_B=\frac{8\pi\alpha s}{q^2}N_e N_p,~ N_e=\frac{1}{s}\bar{u}(p_1')\hat{p} u(p_1);~
N_p=\frac{1}{s}\bar{u}(p')\Gamma_\nu(q) u(p)p_1^\nu.
\label{eq:eqb}
\ee
The phase volume of the final state
\be
d\Gamma=\frac{d^3p_1'}{2E_1'}\frac{d^3p'}{2E'}(2\pi)^{-2}\delta^4(p_1+p-p_1'-p')
\ee
can also be simplified, in peripheral kinematics. Let us introduce an auxiliary  integration over the transferred
momentum as $\int d^4q\delta^4(p_1-q-p_1')=1$, and use the
Sudakov's parameterization:
\be
q=\alpha \tilde{p}+\beta p_1+q_\bot, \tilde{p}=p-\frac{M^2}{s}p_1, q_\bot p_1=q_\bot p=0,
d^4q=\frac{s}{2}d\alpha d\beta, q^2\approx -\vec{q}^2<0.
\label{eq:eqn}
\ee
Performing the integrations on the variables $\alpha$, $\beta$ by means of $\delta$ functions corresponding to the on mass shell conditions of the final electron and proton:
\be
\int d\alpha d^4p_1'\delta^4(p-q-p_1')\delta((p-q)^2-m^2)=\frac{1}{s},
\ee
and a similar expression for the scattered proton, we obtain
\be
d\Gamma=\frac{d^2q_\bot}{2s}\frac{1}{(2\pi)^2}.
\ee
The differential cross section has the form (we omit the subscript '$\bot$')
\be
\frac{d^2\sigma^{ep\to ep}_B}{d^2q}=\frac{4\alpha^2}{(q^2)^2}[F_1^2(q^2)+\tau F_2^2(q^2)], \tau=\frac{\vec{q}^2}{4M^2}.
\ee
Let us consider now the $s$-channel discontinuity of the forward scattering
TPE amplitude, summed over the spin states, with an electron and a proton in the intermediate state. Using the Cutkovsky rules, it can be written
in the form
\ba
A&=&\Delta_s \sum A(s,t=0)=\frac{(4\pi\alpha)^2}{(2\pi)^4}\int\frac{d^4q}{(q^2)^2}(2\pi i)^2
\delta((p_1-q)^2-m^2)\delta((p+q)^2-M^2) \nn \\
&&Tr \hat{p}_1\gamma_\mu(\hat{p_1}-q)\gamma_\nu
Tr(\hat{p}+M)\Gamma_{\mu_1}(q)(\hat{p}+\hat{q}+M)\Gamma_{\nu_1}(-q)
g^{\mu\mu_1}g^{\gamma_\nu\gamma_{\nu_1}}.
\label{eq:eq9}
\ea
Applying the expression (\ref{eq:eqa}) for the photon Green function we obtain for the differential
distribution with respect to the transferred momentum:
\ba
\left .\frac{d^2 A}{d^2q}\right |_{pole}=\frac{32s\alpha^2}{(q^2)^2}[F_1^2(q^2)+\tau F_2^2(q^2)],
\ea
which verifies the optical theorem:
\ba
\left . \frac{d^2A}{d^2q}\right |_{pole}=8s\frac{d^2\sigma_B}{d^2q}.
\label{eq:eq11}
\ea

The cross section for inelastic scattering $ep\to eX$ can be obtained writing the matrix element in the form
\ba
M_{inel}=\frac{8\pi\alpha s}{q^2}N_e N_X, N_X=\frac{1}{s}p_1^\mu J^X_\mu,
\ea
where $J^X_\mu$ is the current describing the the transition from the initial proton to the state $X$ when
interacting with the electromagnetic field. Applying the condition of current conservation:
\ba
q_\mu J^X_\mu\approx (\beta p_1+q_\bot)_\mu J^X_\mu=0,
\label{eq:eqc}
\ea
we find
\ba
N_X=|\vec{q}|\frac{1}{s_2}(\vec{e}\vec{J}^X), ~\vec{e}=\frac{\vec{q}}{|\vec{q}|},
\ea
where  $(p+q)^2=M^2-\vec{q}^2+s_2$ is the invariant mass squared of the proton block,  with $s_2=2qp$ and
$\vec{e}$ is the polarization vector of the virtual photon.

The quantity $\vec{e}\vec{J}^X$ can be expressed in terms of the photo-production cross section:
\ba
\sigma^{\gamma^*p\to X}(s_2)=\frac{\pi\alpha}{2s_2}\int|(\vec{J}^X\vec{e})|^2d\Gamma_X,
\ea
where $d\Gamma_X$ is the phase volume of the created set of particles $X$.
The final expression is:
\ba
\frac{d^2\sigma^{ep\to eX}}{d^2q}=\frac{2\alpha\vec{q}^2}{\pi^2(q^2)^2}
\int\limits_{s_{th}}^s\frac{ds_2}{s_2}\sigma^{\gamma^*p\to X}(s_2,\vec{q}).
\ea
The relevant discontinuity of the forward scattering amplitude can be expressed in terms of the cross section using the optical theorem (\ref{eq:eq11}).

Along the world line, the proton firstly absorbs the virtual photon emitted by
electron (with positive energy) and subsequently, it emits the virtual photon.
The photon absorption precedes the emission of the final state photon.

Therefore the total Compton scattering amplitude {\cal A}, can be written as the sum of an advanced ${\cal A}_{adv}$ and a retarded ${\cal A}_{ret}$ amplitude:
\be
{\cal A}={\cal A}_{ret}+{\cal A}_{adv}.
\label{eq:eql}
\ee
Our approach consists in considering only the retarded part $ {\cal A}_{ret}$.
This amplitude has all the singularities in the region of where the invariant variable $s_2$ is positive: the pole,
situated at $s_2=\vec{q}^2$, which corresponds to a single proton intermediate state; and a sequence of
cuts. The first of them is located at $s_2=\vec{q}^2+2Mm_\pi$, and corresponds to a $N\pi$ state. Further cuts, corresponding to more complicate sets of particles, are located at higher values of $s_2$.
More accurate considerations require to take into account the state corresponding to a proton with a nucleon-antinucleon pair. The relevant
threshold starts
at $s_2=-(\vec{q}^2+8M^2)$. In particular, the contribution to the total cross section which takes into account the identity
of two nucleons in the final state is not described by the retarded part of Compton amplitude.

The retarded amplitude has also a left cut, corresponding to a $pp\bar{p}$ state in the crossed ($u$) channel.

\begin{figure}
\begin{center}
\includegraphics[width=12cm]{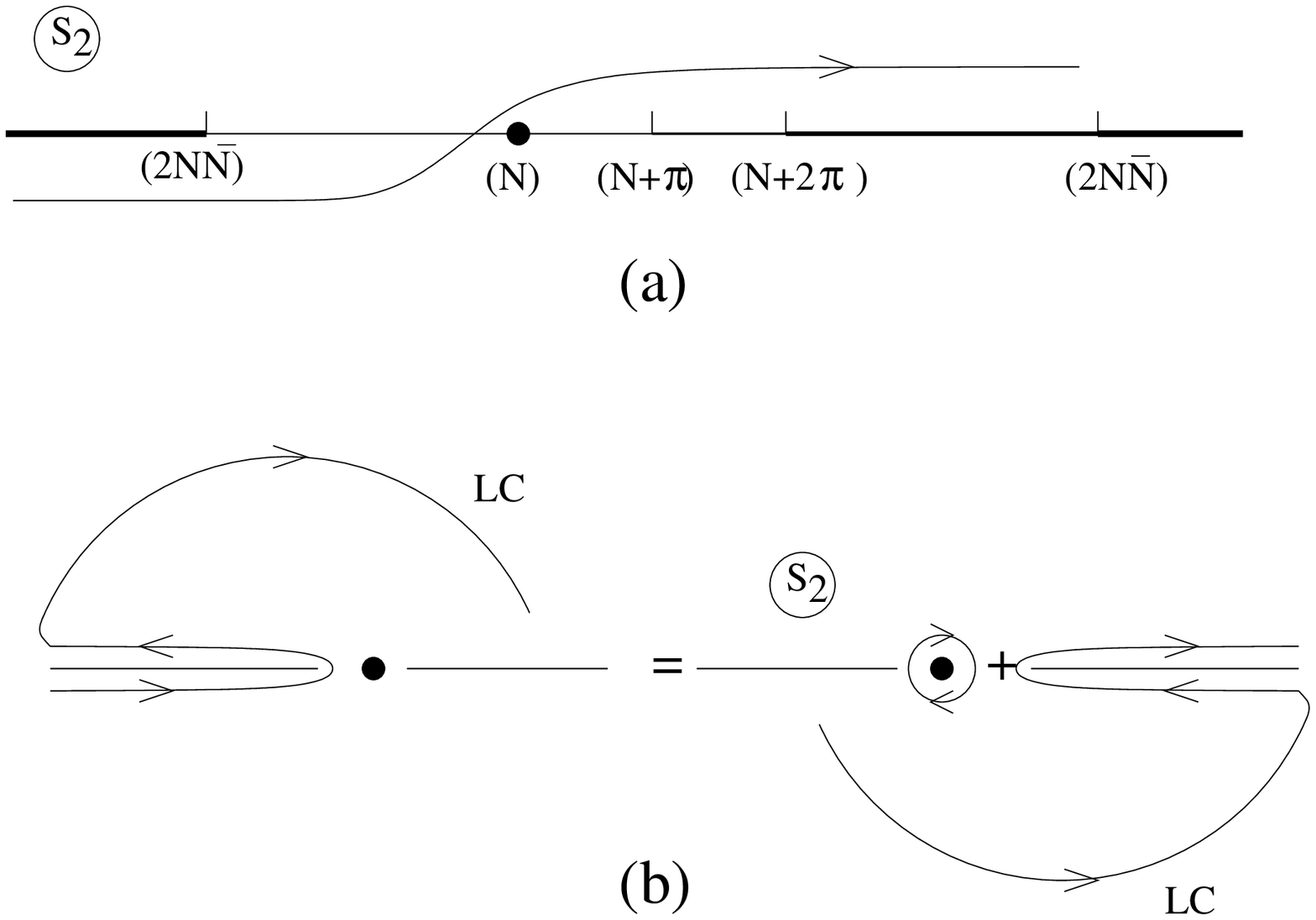}
\caption{\label{Fig:contour} Illustration of singularities along the $s_2$ real axis with the open contour C (a), and with the contour C closed (b). LC stays for large circle contribution.}
\end{center}
\end{figure}

The advanced part of the Compton amplitude has the same kind of
singularities, with the replacement $s_2$ by $u_2=-2pq$. The discontinuities of both parts of the Compton amplitude obey
the current conservation condition. This last statement follows from the fact that the relevant contributions
can be measured in an experiment. The retarded forward Compton scattering amplitude summed over the spin states of the
electron and the proton can be written in the form:
\ba
\frac{d^2{\cal A}_{ret}}{d^2q}=\frac{2^6\pi^2\alpha^2s}{(q^2)^2}g^6\int_C\frac{ds_2}{2\pi i}T,
\ea
where $g$ is the $\pi$ nucleon coupling constant, $g= 13.6\pm 0.3$
\cite{Ti91}.
$T$ is proportional to the light-cone projection of the forward Compton scattering tensor
\ba
T=\frac{1}{4s^2}\sum_{\lambda}\bar{u}^\lambda(p)O^{\mu\nu}u^\lambda(p)p_{1\mu}p_{1\nu}.
\ea
The contour of integration in the $s_2$ plane is located along the real axes in the physical regions of the
$s_2$ and $u_2$ channels (see Fig. \ref{Fig:contour}a):
\be
-\infty-i0<s_2<+i0+\infty.
\ee
 The sum rule appears from the equality of the expression for ${\cal A}$ with the contour closed to the singularities of
the positive part of the $s_2$ real axis and to the singularity of the real axes of $s_2$ plane with  negative values
 of $s_2$ (see Fig. \ref{Fig:contour}b):
 \ba
\frac{d^2\Delta_u {\cal A}_{ret}}{d^2q}=\left .\frac{d^2{\cal A}_{ret}}{d^2q}\right |_{pole}+\frac{d^2\Delta_s {\cal A}_{ret}}{d^2q}.
 \ea
The contribution of the so called "large half circles" tends to zero when the radii of the large circle tends to
infinity. This statement holds due to the gauge invariance property of the discontinuities of the light-cone
projection of the Compton scattering amplitude. With the help of Eq. (\ref{eq:eqc}) one can write:
\be
\frac{1}{s^2}\bar{u}(p)O_{\mu\nu}u(p)p_1^\mu p_1^\nu=\frac{1}{s_2^2}\bar{u}(p)O_{\mu\nu}u(p)q_\bot^\mu q_\bot^\nu.
\ee
\section{Calculation of the left cut contribution}

\begin{figure}
\begin{center}
\includegraphics[width=12cm]{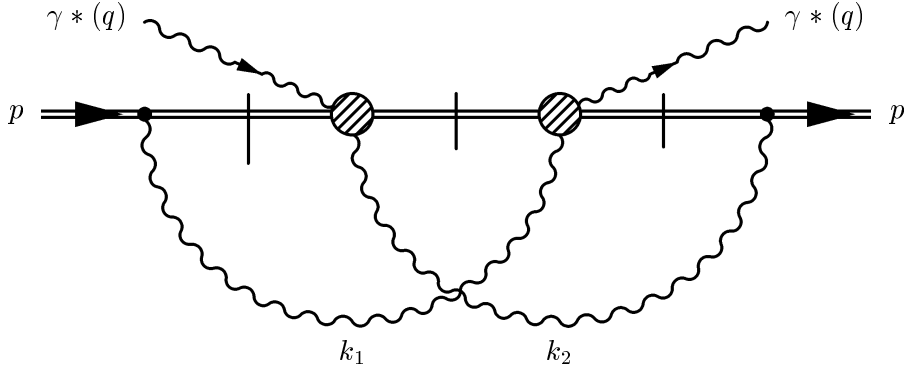}
\caption{\label{Fig:Fig4} Diagram for the amplitude for Compton scattering. }
\end{center}
\end{figure}

The Feynman diagram which corresponds to the left cut contribution is drawn in Fig. \ref{Fig:Fig4}, where the empty circles denote the
amplitudes of sub-processes $p+\gamma^*\to p+\pi_0$ and $p+\pi_0\to p+\gamma$ and the crossed lines represent the on
mass shell protons and the antiproton. One can be convinced that only this type of amplitudes with two protons
and an anti-proton (all particles are on mass shell) corresponds to the nearest left cut, if we neglect possible exotic states. The
intermediate states with $\pi$-mesons give the main numerical contribution. States involving heavier
scalar or pseudo scalar particles as well as vector mesons are suppressed at least by a factor equal to the square of the ratio of their masses.
This follows from the convergence of loop momentum integrals in the region where their values are much lower than the rest energy of the proton.
The left cut contribution does not suffer from infrared as well as ultraviolet  divergences,
and obeys the gauge invariance conditions applied to both exchanged photons.

Sudakov's form of the momenta of the intermediate mesons reads as $k_i=\alpha_i \tilde{p}+\beta_i p_1 +k_{i\bot}$.
We perform the $\beta_{1,2}$ integrations as well as the $s_2$ integration using the on mass shell conditions of the
two protons and the anti-proton:
\ba
&\int& ds_2d^4k_1d^4k_2\delta((p-k_1)^2-M^2)\delta((p-k_2)^2-M^2)\delta((p+q-k_1-k_2)^2-M^2)
\nn\\
&=&\frac{1}{4}\int
\frac{d\alpha_1\theta(c_1)}{c_1}\frac{d\alpha_2\theta(c_2)}{c_2}
\frac{\theta(-c)}{|c|}d^2\vec{k}_1d^2\vec{k}_2=\pi^2\int d\Gamma, \nn \\
c_1&=&1-\alpha_1;c_2=1-\alpha_2; c=1-\alpha_1-\alpha_2.
\label{eq:eqcc}
\ea
We obtain for the $u$ channel discontinuity of the forward scattering amplitude  summed on spin states:
\be
\left (\frac{d^2\Delta {\cal A}_{ret}}{d^2 q}\right )_{left}=s\frac{\alpha^2g^4}{\pi^4(q^2)^2}\Phi(\vec{q}),~ \mbox{with~}
\Phi(\vec{q})=\int d\Gamma \frac{T}{(k_1^2-m^2)(k_2^2-m^2)},
\label{eq:Phi}
\ee
where $m$ is the pion mass; the quantity $T$
has the form :
\ba
T&=&\frac{1}{4s^2}Tr(\hat{p}+M)(\hat{p}-\hat{k}_2-M)
[A_{11}s+A_{12}\hat{p}_1\hat{q}](\hat{p}+\hat{q}-\hat{k}_1-\hat{k}_2+M) \nn \\
&&[A_{21}s+A_{22}\hat{q}\hat{p}_1](\hat{p}-\hat{k}_1-M).
\label{eq:eqT}
\ea
The details are given in the Appendix.
Using the expressions given in Appendix one can be convinced that the quantities in the square brackets
in the expression for $T$, Eq. (\ref{eq:eqT}), which are the light-cone projections of the amplitudes for the subprocesses $p+\gamma^*\to p+\pi_0$ and
$p+\pi_0\to p+\gamma$ vanish in the limit $\vec{q}\to 0$. This property is a consequence of gauge invariance.

The behavior of the function $\Phi(\vec{q})$ (Eq. \ref{eq:Phi}) averaged over the azimuthal
angle $\varphi_q$:  $\bar{\Phi}(z)=
\int d\varphi_q/(2\pi)\Phi(\vec{q})$, with $z=\vec{q}^2/M^2$ and  $d^2q=(d\vec{q}^2d\varphi_q)/2$ is presented
in Fig. \ref{Fig:res}.
\section{Discussion and conclusion}
We note that the quantity $T$, Eq. (\ref{eq:eqT}), is dimensionless. Keeping in mind its gauge properties we conclude that
$T\sim |\vec{q}|^2/M^2$. This property provide the ultra-violet convergence when integrating on ${k}_{1,2}$. Moreover
its characteristic value is much lower than the proton mass, which follows from the analysis of denominators $D_i$, Eq. (\ref{eq:eqdd}).

\begin{figure}
\begin{center}
\includegraphics[width=12cm]{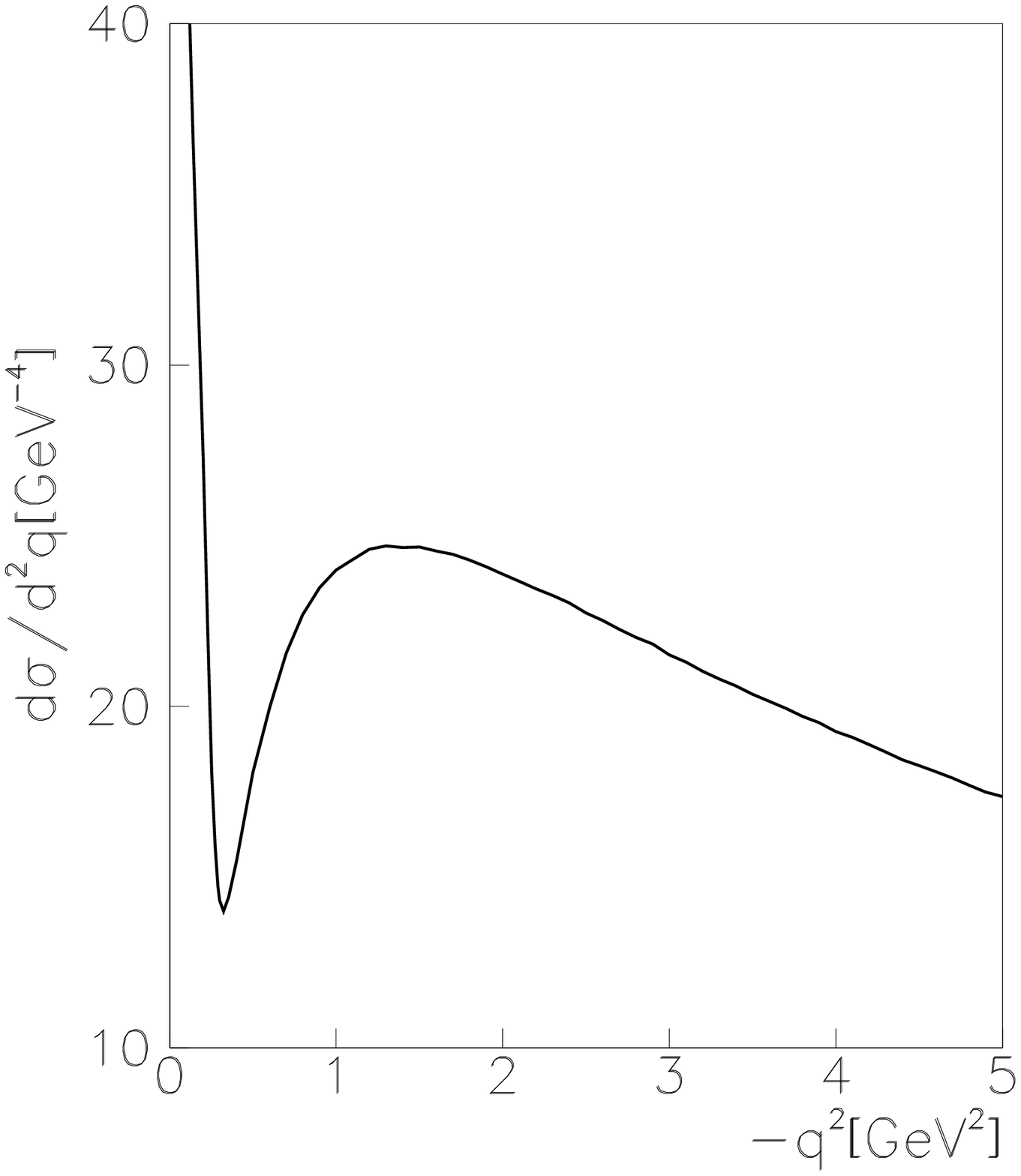}
\caption{\label{Fig:sum:cross} Prediction for $\displaystyle\frac{d\sigma^{ep\to eX}}{d\vec{q}^{\,\,\,2}}$ in GeV$^{-4}$ (see Eq. (\protect\ref{eq:sumrules})). }
\end{center}
\end{figure}

\begin{figure}
\begin{center}
\includegraphics[width=12cm]{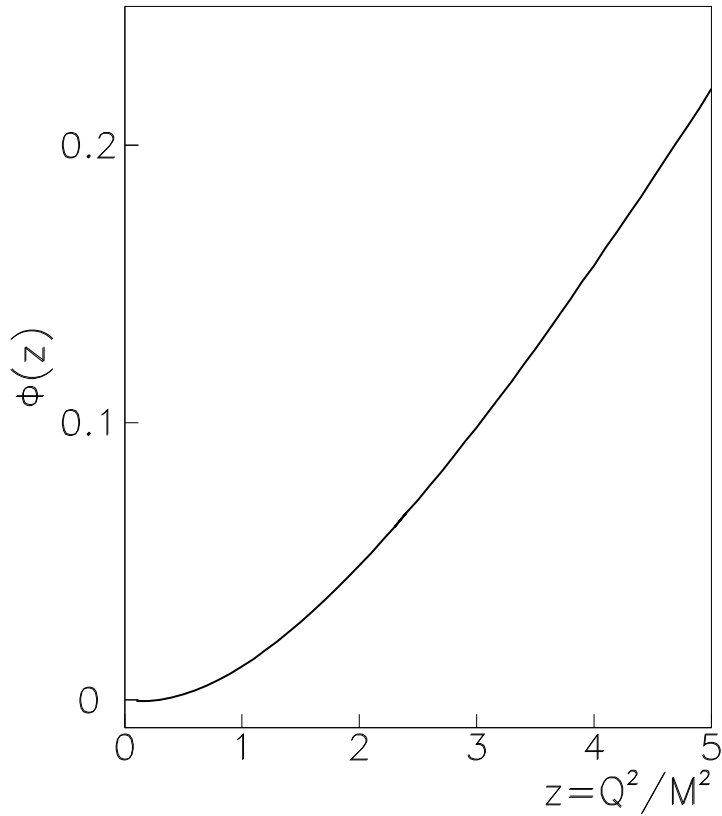}
\caption{\label{Fig:res} Behavior of the function $\bar \Phi(z)$ (see Eq. (\protect\ref{eq:Phi})). }
\end{center}
\end{figure}

For the case of large angles scattering the suppression factor of the left cut contribution is  $\vec{q}^2/s<<1$.
So it can be interpreted as higher twists contributions. Therefore it can be neglected when considering the
interference of Born and Box type Feynman amplitudes for TPE in large angles electron (positron) scattering on a proton
\cite{epodd}.
The sum rule obtained above can be formulated in terms of cross sections. It has the form:
\be
\label{eq:sumrules}
 \frac{g^4}{8\pi^4}\bar{\Phi}\left (\frac{\vec{q}^2}{M^2}\right )=-4\left [F_1^2(\vec{q}^2)+\frac{\vec{q}^2}{4M^2}F_2^2(\vec{q}^2)\right ]+
(\vec{q}^2)^2\frac{d\sigma^{ep\to eX}}{d\vec{q}^{\,\,\,2}}.
\ee
In Fig. \ref{Fig:sum:cross}  the prediction for the cross section of ${ep\to eX}$
assuming dipole form-factor behavior is shown.

Calculating the derivative of both sides on $\vec{q}^2$ at $\vec{q}^2=0$ and using the expressions for the charge radii and the
anomalous magnetic moment of proton in terms of its form factors
\ba
F_1'(0)=\frac{1}{6}<r_p^2>,~F_2(0)=\mu,
\ea
we obtain
\ba
\label{eq:equality}
 \frac{g^4}{8\pi^4}\bar{\Phi}'(0)=-4\left [\frac{1}{3}<r^2_p>+\frac{\mu^2}{4M^2}\right ]+
\frac{2}{\pi^2\alpha}\int\limits_{s_{th}}^s\frac{ds_2}{s_2}\sigma_{tot}^{\gamma p\to X}(s_2).
\ea
For the case of moderately high energies the upper limit, $s$,  of the integration over $s_2$ can be replaced by $\infty$, due to the
absence of Pomeron contribution to the photo-production cross section $\sigma^{\gamma p\to X}$.

Using the known values \cite{PdG} of $\sqrt{<r^2_p>}=4.35/M $, $\mu=1.79 $ and substituting the value of the
derivative $M^2\bar{\Phi}'(0)=-2.9*10^{-3}$ (see details in Appendix),
we obtain for the weighted integral of cross section of photoproduction for the case of moderate high energies:
\ba
\int\limits_{s_{th}=0.28 GeV^2}^{s=8 GeV^2}\frac{d s}{s}\sigma^{\gamma p\to X}(s)=\frac{0.9}{M^2}.
\ea
We imply that the total photoproduction cross section $\sigma^{\gamma p\to X}(s)$ is a decreasing function of $s$.
Due to this reason we use the results on the total cross section for Ref. \cite{PdG} in a restricted region of $s$
($s<8$ GeV$^2$), as data on the separate contribution of definite
channels are absent. For numerical estimation of (\ref{eq:equality}) we obtain:
\ba
-0.13\frac{1}{M^2}\approx-26.0\frac{1}{M^2}+25.0\frac{1}{M^2},
\ea
which is in reasonable agreement with the sum rules statement. For the case of ultra high energies the Pomeron
contribution becomes essential. Nevertheless taking into account the universal character of its interaction with
hadrons, a sum rule can still be derived through a linear combination involving proton and neutron, built in such a way to eliminate Pomeron contribution \cite{dubn}.

The region of validity of our formulas is determined by peripheral kinematics
\ba
\frac{Q^2}{s}<<1, Q^2\sim M^2.
\ea
The accuracy of the formulas given above is determined by the omitted terms
\ba
1+O\left (\frac{M^2}{s},\frac{Q^2}{s}\right ).
\ea
Furthermore, we do not consider the pure QED radiative corrections, as their contribution does not exceed a few percents.

\section{Acknowledgments}
Two of us (V.V.B and E.A.K.) are grateful to CEA Saclay for warm  hospitality,  to INTAS grant 05-1000-008-8323  and to grant MK-2952.2006.2 for financial support.

\section{Appendix}

Solving the on-mass shell condition for the protons, in frame of the Sudakov parameterization, we can express all the
kinematic invariants of the problem in terms of
euclidean two-vectors and energy fractions $\alpha_{1,2}$:
\be
s\beta_1=-\frac{\vec{k}_1^2+\alpha_1M^2}{1-\alpha_1};k_1^2=-\frac{\vec{k}_1^2+M^2\alpha_1^2}{1-\alpha_1};
s\beta_2=-\frac{\vec{k}_2^2+\alpha_2M^2}{1-\alpha_2};
k_2^2=-\frac{\vec{k}_2^2+M^2\alpha_2^2}{1-\alpha_2},
\ee
with $c$, $c_{1,2}$ defined in Eq. (\ref{eq:eqcc}).
>From  the on mass shell condition for the anti-proton we obtain
\be
s_2=\frac{d_q}{cc_1c_2}, d_q=M^2\alpha_1\alpha_2(2-\alpha_1-\alpha_2)+c_1c_2[\vec{q}^2-2\vec{q}(\vec{k}_1+\vec{k}_2)+
2\vec{k}_1\vec{k}_2]+\alpha_1c_1\vec{k}_2^2+\alpha_2c_2\vec{k}_1^2.
\ee
Keeping in mind that the region of the main contribution corresponds to small $|\vec{k}_{1,2}|$ and $\alpha_1\approx
\alpha_2\approx (-c)\approx 2/3$,  one sees that  $s_2<-8M^2$.
Applying the on-mass-shell condition, the light-cone projection of the amplitude of the subprocess $\gamma^*(q)+p(p_1-k_1)\to \pi^0(k_2)+p(p+q-k_1-k_2)$
can be written in the form:
\ba
\bar{u}(p+q-k_1-k_2)\biggl[
\gamma_5\frac{\hat{p}+\hat{q}-\hat{k}_1+M}{D_2}p_1+p_1\frac{\hat{p}-\hat{k}_1-\hat{k}_2+M}{D}\gamma_5
\biggr]
u(p-k_1)
\nn \\
\approx
\bar{u}(p+q-k_1-k_2)\biggl[
s\gamma_5\left (\frac{c_1}{D_2}+\frac{c}{D}\right )+\gamma_5\hat{q}\hat{p}_1
\left (\frac{1}{D_2}+\frac{1}{D}
\right )
\biggr]u(p-k_1).
\ea
The expressions for $D_i$ are
\ba
D_1&=&(p+q-k_2)^2-M^2=\frac{d_q}{c_1c},~D_2=(p+q-k_1)^2-M^2=\frac{d_q}{c_2c}; \nn \\
D&=&(p-k_2-k_1)^2-M^2=-\frac{d}{c_2c_1},2k_1k_2=D; \nn \\
d&=&M^2\alpha_1\alpha_2(1+c)+2c_2c_1\vec{k}_1\vec{k}_2+\alpha_1c_1\vec{k}_2^2
+\alpha_2c_2\vec{k}_1^2.
\label{eq:eqdd}
\ea
Writing the quantity $T$ as
\be
T=A_{11}A_{21} C_{1121}+A_{11}A_{22} C_{1122}+A_{12}A_{21} C_{1221}+A_{12}A_{22} C_{1122},
\ee
we find
\be
A_{11}=A_{21}=cc_1c_2\left [\frac{1}{d_q}-\frac{1}{d}\right ];
~A_{12}=c_1\left [\frac{c}{d_q}-\frac{c_2}{d}\right ];
~A_{22}=c_2\left [\frac{c}{d_q}-\frac{c_1}{d}\right ].
\ee
The coefficients  have the  form:
\ba
C_{1222}&=&\frac{c\vec{q}^2}{4}d_0; ~
C_{1121}=\frac{1}{2}\left [(\vec{q}\vec{k}_1)(\vec{q}\vec{k}_2)-(k_1^2+\vec{q}\vec{k}_1)(k_2^2+\vec{q}\vec{k}_2)+
\frac{s_2}{2}d_0 \right ]; \nn \\
C_{1221}&=&\frac{1}{2}k_2^2[\alpha_1\vec{q}\vec{k}_2+c_2\vec{q}\vec{k}_1-\frac{1}{2}\alpha_1\vec{q}^2]-\vec{q}\vec{k}_2k_1k_2-\frac{1}{4}\alpha_2\vec{q}^2k_1^2; \nn \\
C_{1122}&=&\frac{1}{2}k_1^2[\alpha_2\vec{q}\vec{k}_1+c_1\vec{q}\vec{k}_2-\frac{1}{2}\alpha_2\vec{q}^2]-\vec{q}\vec{k}_1k_1k_2-\frac{1}{4}\alpha_1\vec{q}^2k_2^2, \nn \\
d_0&=&d_0=k_1^2\alpha_2+k_2^2\alpha_1-2k_1k_2.
\ea
The calculation of the derivative of $\Phi(\vec{q}^2)$ at $\vec{q}^2=0$ gives:
\be
 M^2\Phi'(0)=M^2\int\frac{d^2k_1}{2\pi}\frac{d^2k_2}
 {2\pi}\frac{1}{\Lambda_1+m^2c_1}\frac{1}{\Lambda_2+m^2c_2}\int\limits_0^1
 d\alpha_1\int\limits_{c_1}^1d\alpha_2\frac{1}{2d^2}[A_1+A_2+A_3+A_4],
\ee
with
\ba A_1&=&\frac{c_1^2c_2^2}{d^2}(\vec{k}_1+\vec{k}_2)^2[-2c_1c_2c\Lambda_1\Lambda_2+d^2-d(\alpha_1c_1\Lambda_2+\alpha_2c_2\Lambda_1)];
\nn \\
A_2&=&-2\alpha_1\alpha_2[\alpha_1c_1\Lambda_2+\alpha_2c_2\lambda_1-d]; \nn \\ A_3&=&\frac{c_1c_2^2}{d}(\vec{k}_1+\vec{k}_2)[c_2\Lambda_1(\alpha_2\vec{k}_1+c_1\vec{k}_2)-d\vec{k}_1]; \nn \\ A_4&=&\frac{c_1^2c_2}{d}(\vec{k}_1+\vec{k}_2)[c_1\Lambda_2(\alpha_1\vec{k}_2+c_2\vec{k}_1)-d\vec{k}_2]; \nn \\
\Lambda_1&=&\vec{k}_1^2+M^2\alpha_1^2+m^2c_1;\Lambda_2=\vec{k}_2^2+M^2\alpha_2^2+m^2c_2.
\ea

\end{document}